\documentclass[review]{elsarticle}

\usepackage{hyperref}
\journal{Physica A}

\begin{document}

\begin{frontmatter}

\title{Role of inhibitory neurons in temporal correlations of critical and supercritical spontaneous activity}

\author{Dario Raimo}

\address{Dept. of Physics "Ettore Pancini", University of Naples "Federico II", 80100 Napoli, Italy}

\author{Alessandro Sarracino, Lucilla de Arcangelis}

\address{Dept. of Engineering, University of Campania "Luigi Vanvitelli", 81031 Aversa (CE), Italy}

\begin{abstract}
Experimental and numerical results suggest that the brain can be viewed as a system acting close to a critical point, as confirmed by scale-free distributions of relevant quantities in a variety of different systems and models. Less attention has received the investigation of the temporal correlation functions in brain activity in different, healthy and pathological, conditions. Here we perform this analysis by means of a model with short and long-term plasticity which implements the novel feature of different recovery rates for excitatory and inhibitory neurons, found experimentally. We evidence the important role played  by inhibitory neurons in the supercritical state: We detect an unexpected oscillatory behaviour of the correlation decay, whose frequency depends on the fraction of inhibitory neurons and their connectivity degree. This behaviour can be rationalized by the observation that bursts in activity become more frequent and with a smaller amplitude as inhibition becomes more relevant. 
\end{abstract}
 

\end{frontmatter}

%\linenumbers

\section{Introduction}
 
Spontaneous brain activity has a characteristic spatio-temporal pattern, critical avalanches, first observed in organotypic cultures from coronal slices of rat cortex \cite{beggsPlenz2003}. The distribution of avalanche sizes consistently exhibits the scaling $P(s)\propto s^{-1.5}$. Critical avalanche dynamics is also found in cortical activity of awake moneys \cite{petermann2009spontaneous} as well as in human fMRI (functional Magnetic Resonance Imaging) \cite{fransson2012early} and MEG (MagnetoEncephaloGraphy) recordings \cite{shriki2013}. However, the existence of a different universality class has been also proposed in the literature \cite{font19}.
Inspired in self-organized criticality (SOC) \cite{BTW}, a model implementing the main physiological features of real neurons \cite{lucilla2006,lucilla2012,lucillaActD} was able to reproduce the experimental MEG spectra and avalanche statistics. Neuronal network models implementing both short- and long-term synaptic plasticity also reproduce neuronal avalanche behaviour at a tunable parameter critical value \cite{levina,van2018critical}: The avalanche size distribution exhibits a power law behaviour up to avalanche sizes as large as the whole system in the critical state, whereas an exponential cutoff sets in at smaller avalanches in the subcritical state and a bimodal behaviour, with an excess of large avalanches, is observed in the supercritical state. Alternative approaches describe brain activity by phase models \cite{finger}.
The hypothesis that the brain acts as a system close to a critical point has inspired the implementation of statistical models on the brain connectome \cite{Nuzzi} and leads to the optimization of the dynamic range \cite{kinouchi2006optimal} and the tremendous variability in neuronal activity patterns observed in the brain at all scales. At the same time, deviations from criticality are observed in systems with pathological behaviour \cite{critrev}.

Inhibitory neurons have a crucial role in the organization of brain activity \cite{lombardiWaiting,lombchaos,poil,dalpo19,gs20}. Experiments have shown that they are functional hubs, able to orchestrate synchrony in developing hippocampal networks \cite{bonifazi2009gabaergic}. Moreover, the percentage (20-30\%) of inhibitory neurons observed in mammal brains has been numerically found to optimize the learning performances since it guarantees, at the same time, the network excitability necessary to express the response and the variability required to confine
the employment of resources \cite{vittorioLearning}. Supercritical conditions can be induced  pharmacologically in vitro and in vivo by hindering inhibitory activity, which suggests that inhibitory neurons have a crucial role in healthy brain activity.

Here we address the question of the role of inhibition in spontaneous activity in different states, critical and supercritical, which correspond to healthy and pathological conditions. We analyse in particular the features of the temporal auto-correlation function in the two states, detected from the properties of the avalanche size distribution, and monitor the dependence of its behaviour on the relevance of inhibitory neurons.

This manuscript has been written in memory of Dietrich Stauffer, who has left to one of us the unconditional belief in science and the wisdom of generosity as precious heritage. 

\section{Neuronal model with short and long-term plasticity}
We consider a network of $N$ neurons, placed randomly in a cube of side $L$, with a fraction $p_{in}$ of inhibitory neurons. Synaptic connections are directed and each neuron has $k_{out}$ outgoing connections, distributed according to the scale-free distribution found experimentally for functional networks \cite{cecchi}, $n(k_{out}) \propto k_{out}^{-2}$ with $k_{out} \in [2,100]$. Since inhibitory neurons are hubs in the network \cite{bonifazi2009gabaergic}, they are assigned among neurons with large out-degree, i.e. $k_{out}\in [k_{min},100]$. The probability to establish a connection depends on the spatial distance $r$ between two neurons according to $p(r) = \frac{1}{r_0} e^{-r/r_0}$ \cite{roerig2002relationships} with $r_0$ a characteristic connectivity range. Initial synaptic strengths $\omega_{ij}$ are chosen to be uniformly distributed within the interval $[0.04,0.06]$. Incoming connections are identified a posteriori after outgoing connections are established. 
Every neuron $i$ is characterized by a membrane potential $v_i$, with random initial value in $[0.5,1[$. At each time step all neurons with a potential exceeding a threshold value $v_i \geq v_c = 1.0$ fire, causing a change in the potential of the pre-synaptic and post-synaptic neurons $j$ as \cite{van2018critical}

$$
v_j(t+1) = v_j(t) \pm v_i(t)   u_i \omega_{ij} \nonumber$$
$$u_{i}(t+1) = u_{i}(t)(1-\delta u) \eqno(1)$$
$$v_i(t+1) = 0 \nonumber
$$

\noindent
where $u_{i}$ is the amount of releasable neurotransmitter, which initially is set equal to one for all synapses of each neuron $i$. The second equation implements the short-term plasticity in the model: The amount of releasable neurotransmitter reduces by roughly $5\%$ after a spike \cite{ikeda2009counting, rizzoli2005synaptic}, therefore $\delta u = 0.05$. The plus or minus sign in Eq.(1) stands for excitatory or inhibitory pre-synaptic neuron, respectively. After a neuron fires, it remains in a refractory state for one time step during which it is unable to receive or send any signal. 

 The recovery of synapses takes a time of the order of seconds \cite{tsodyks1997neural} whereas synaptic transmission is of the order of milliseconds. Therefore, recovery of
 available neurotransmitter is implemented at the end of each avalanche, i.e. all  $u_i$ are increased by an amount $\delta u_{rec}$, which determines if the network will be in a subcritical, critical or supercritical state. Experimental results have evidenced \cite{recov} that the vesicular GABA uptake rate is 5 or 6 times slower than the glutamate uptake; we therefore consider also the case of different recovery rates, $\delta u_{recE}$ and $\delta u_{recI}$, for excitatory and inhibitory neurons, with $\delta u_{recE}=5\delta u_{recI}$, choice that still allows the tuning of a single parameter. At the end of each avalanche, activity is kept ongoing by setting at threshold a random neuron. 

The synaptic network is dynamically modified by long-term plasticity: Whenever a neuron $i$ fires, all weights of outgoing synapses, connecting $i$ to an active post-synaptic neuron $j$ are strengthened according to the voltage variation induced in the post-synaptic neuron
$$
\omega_{ij}(t+1) = \omega_{ij}(t) + \delta \omega_{ij} \eqno(2)
$$
where $\delta \omega_{ij}=\alpha (v_j(t+1)-v_j(t))$ and $\alpha=0.006$ determines the rate of this plastic change. When an avalanche ends, all strengths $\omega_{ij}$ are decreased by the average increase in synaptic strength per bond

$$\omega_{ij} = \omega_{ij}-\sum \delta \omega_{ij}/N_s \eqno(3)$$ 
 
 where $N_s$ is the total number of synapses.
 Synaptic plasticity weakens loops in the network structure \cite{van2016synaptic}. Plastic adaptation is implemented for a fixed number of external stimulations, or until a first synapse is pruned, with the duration of plastic adaptation representing the age or experience of the system. Once the synaptic strengths are sculpted by the activity, we keep the $\omega_{ij}$ fixed and stimulate a further sequence of avalanches by solely implementing short-term plasticity. Data are  averaged over 8000 configurations of networks with $N=5000$ neurons in a cube of side $L=100$ for a connectivity range $r_0 = 7.5$.

\section{Avalanche activity}
By tuning the synaptic recovery parameter $\delta u_{rec}$, it is possible to set the system in different activity states \cite{van2018critical}: subcritical, critical and supercritical. In Fig.1 we show the distribution of avalanche sizes, measured as the number of firing neurons, in the critical state. The distribution exhibits the power law behaviour $s^{-1.5}$ found experimentally for different percentage of inhibitory neurons, where the cut-off at large size decreases for increasing $p_{in}$, since the dissipative role of inhibition hinders the occurrence of large avalanches. The scaling behaviour is quite robust and is recovered for both version of the model, in particular also if we implement different recovery rates for excitatory and inhibitory neurons. In this case, the slow recovery of inhibitory neurons reduces their activity leading to a larger cut-off in the distribution. The same behaviour if detected in the subcritical regime, with a reduced scaling regime and a cut-off at smaller avalanche sizes.

\begin{figure}[htbp]
	\centering
	\includegraphics[width=0.49\textwidth,clip=true]{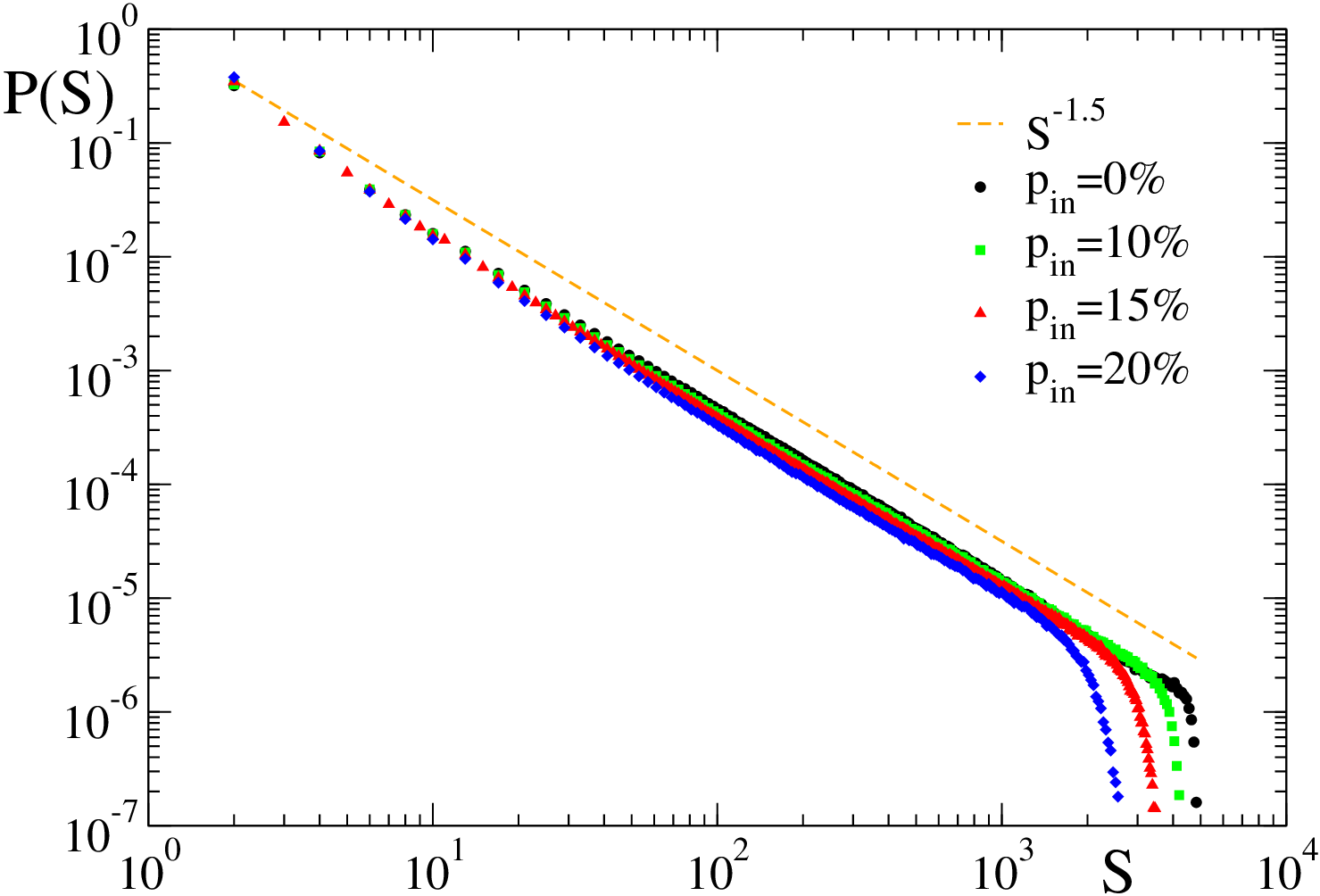}
    \includegraphics[width=0.49\textwidth,clip=true]{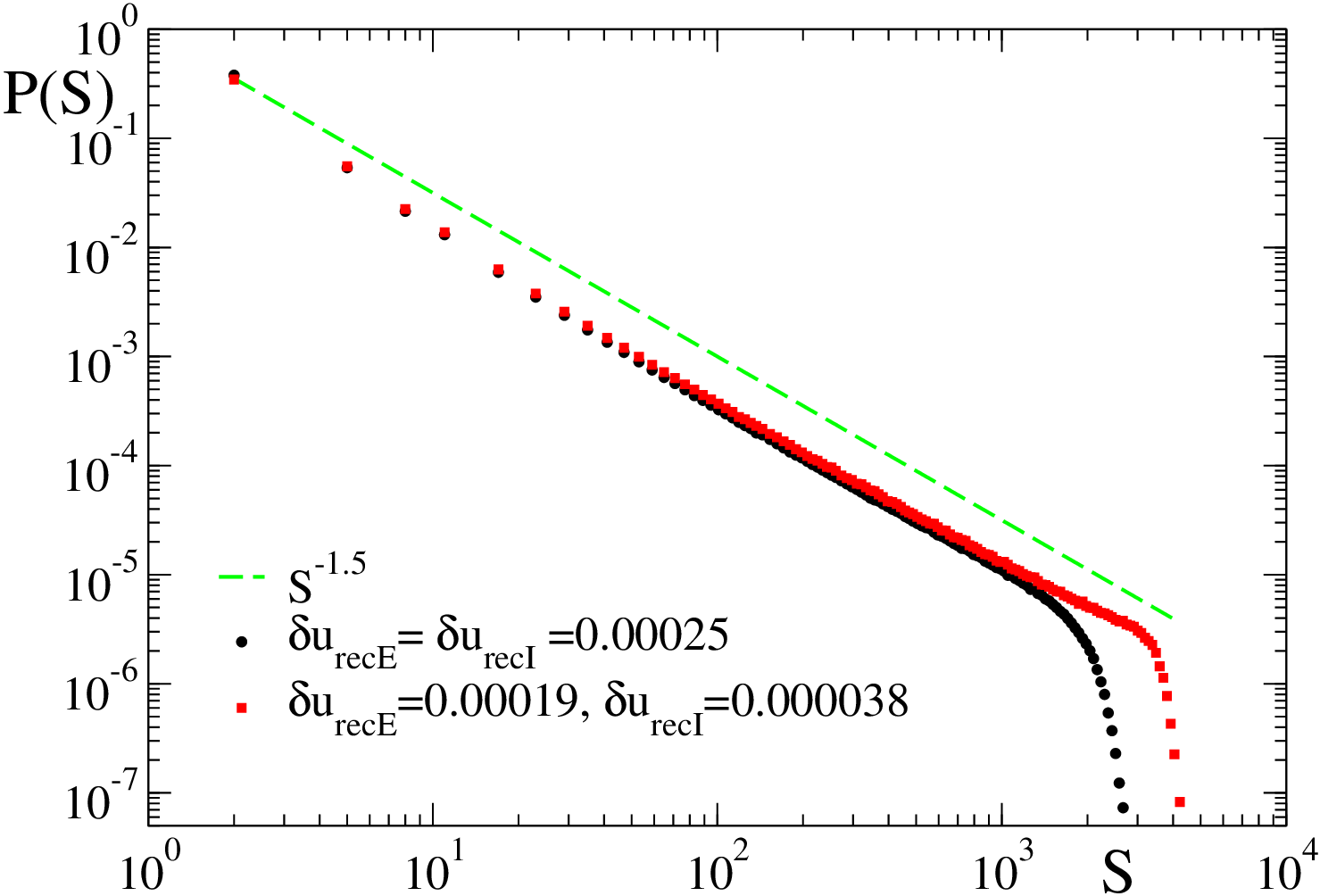}
	\caption{Avalanche size distribution for 8000 configurations of a network of $N=5000$ neurons at criticality: $Left$ for a single value of $\delta u_{rec}$ and for different percentages of inhibitory neurons; $Right$ for $p_{in}=20\%$ and for the two cases of  single-valued and different $\delta u_{rec}$ for excitatory and inhibitory neurons.}
	\label{fig1}
\end{figure}

In the supercritical state the distribution shows a clear bimodal structure, with a power law behaviour followed by a bump at large sizes (Fig.2). This excess of large avalanches is due to an unbalance in activity of excitatory and inhibitory neurons. Indeed, these have a crucial role since not only their percentage controls the cut-off, as in the critical state, but also their connectivity degree. In fact, by tuning $k_{min}$, we control their influence on the activity of the entire network, which is clearly shown by the bump position which, for increasing $k_{min}$, moves to smaller avalanche sizes and increases in height. As for the critical state, different recovery rates limit the dissipative role of inhibitory neurons, leading to a bump at larger sizes and shifting the cutoff. Since inhibition clearly affects spontaneous avalanche activity, both in the critical and supercritical states, we investigate in more detail its role in the properties of the activity signal and the temporal correlation functions.  

\begin{figure}[htbp]
	\centering
	\includegraphics[width=0.49\textwidth,clip=true]{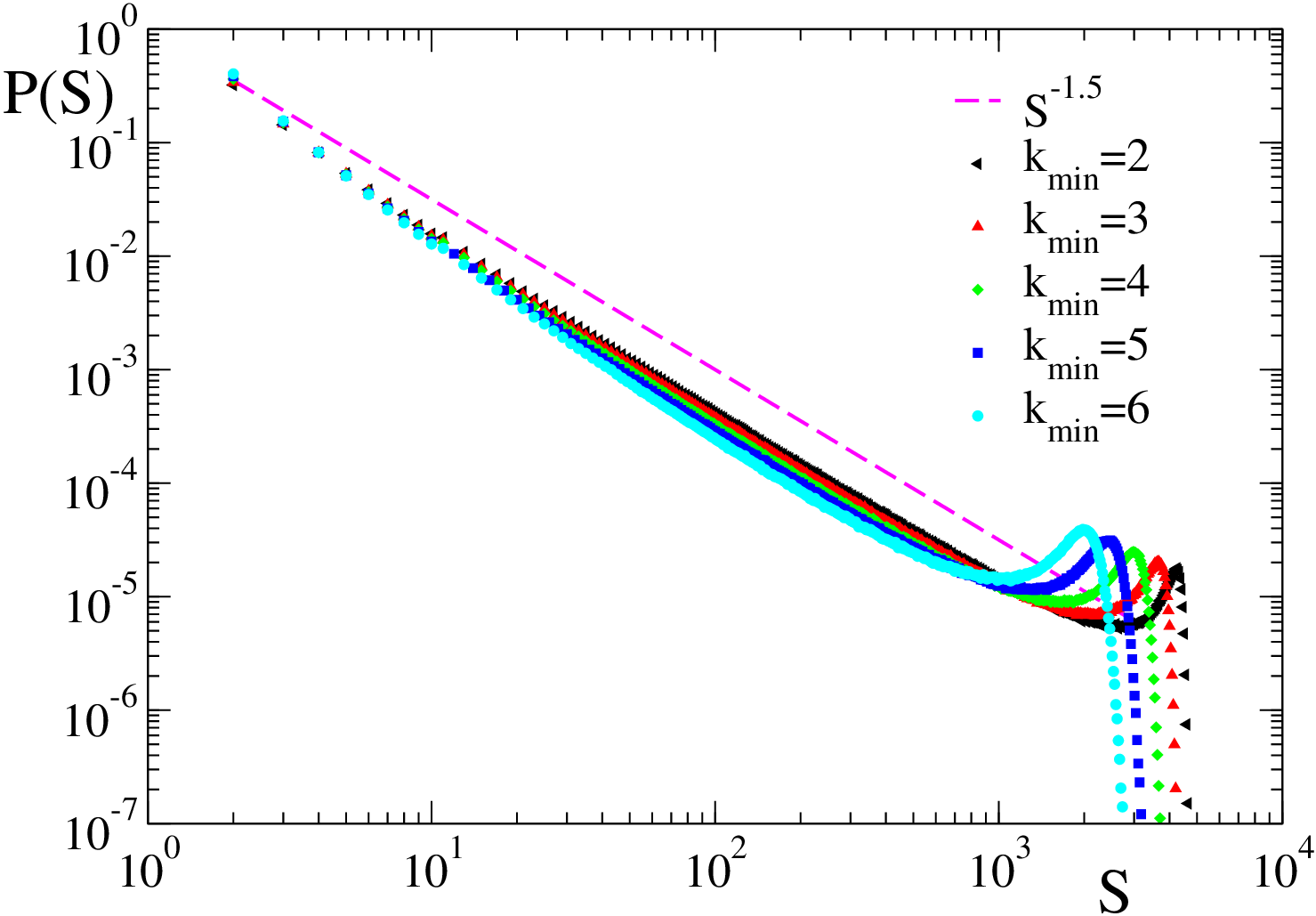}
	\includegraphics[width=0.49\textwidth,clip=true]{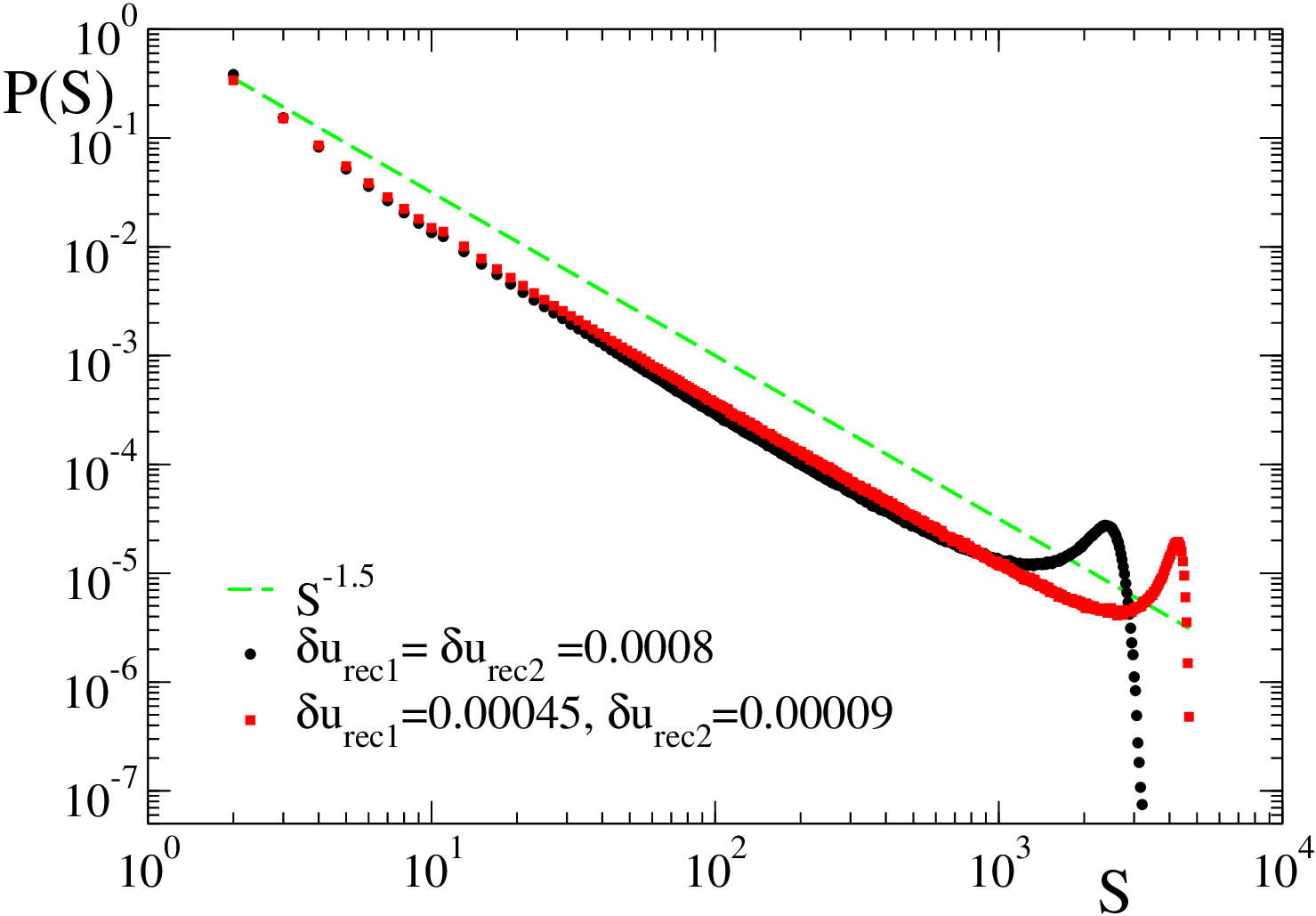}
	\caption{Avalanche size distribution for 8000 configurations of a network of $N=5000$ neurons with $p_{in}=20\%$ in the supercritical state: $Left$ for a single value of $\delta u_{rec}$ and for different minimum connectivity degree of inhibitory neurons; $Right$ for $k_{min}=5$ and the two cases of a single-valued and different $\delta u_{rec}$ for excitatory and inhibitory neurons.}
	\label{fig2}
\end{figure}

\begin{figure}[t]
	\centering
	\includegraphics[width=0.49\textwidth,clip=true]{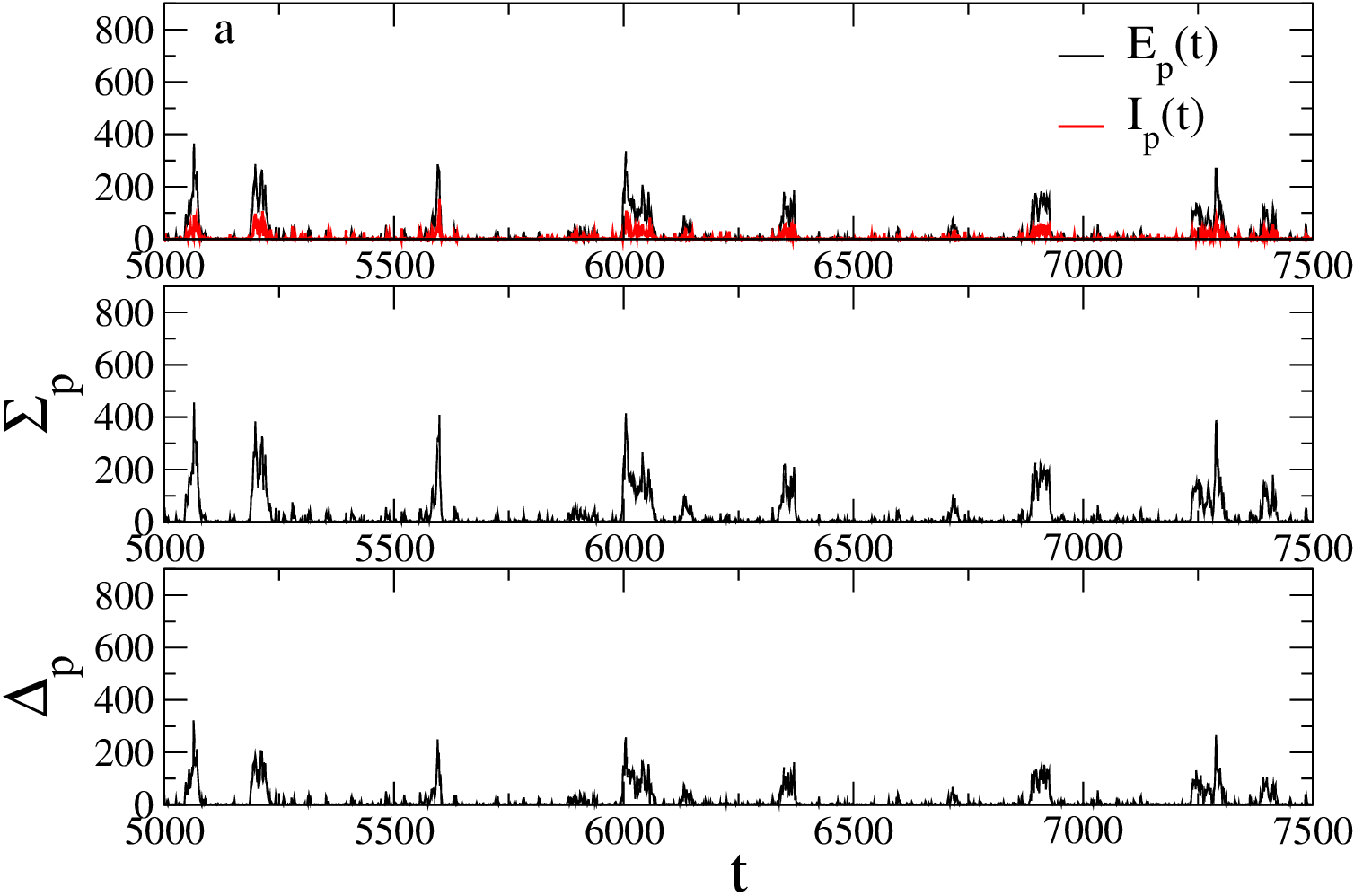}
	\includegraphics[width=0.49\textwidth,clip=true]{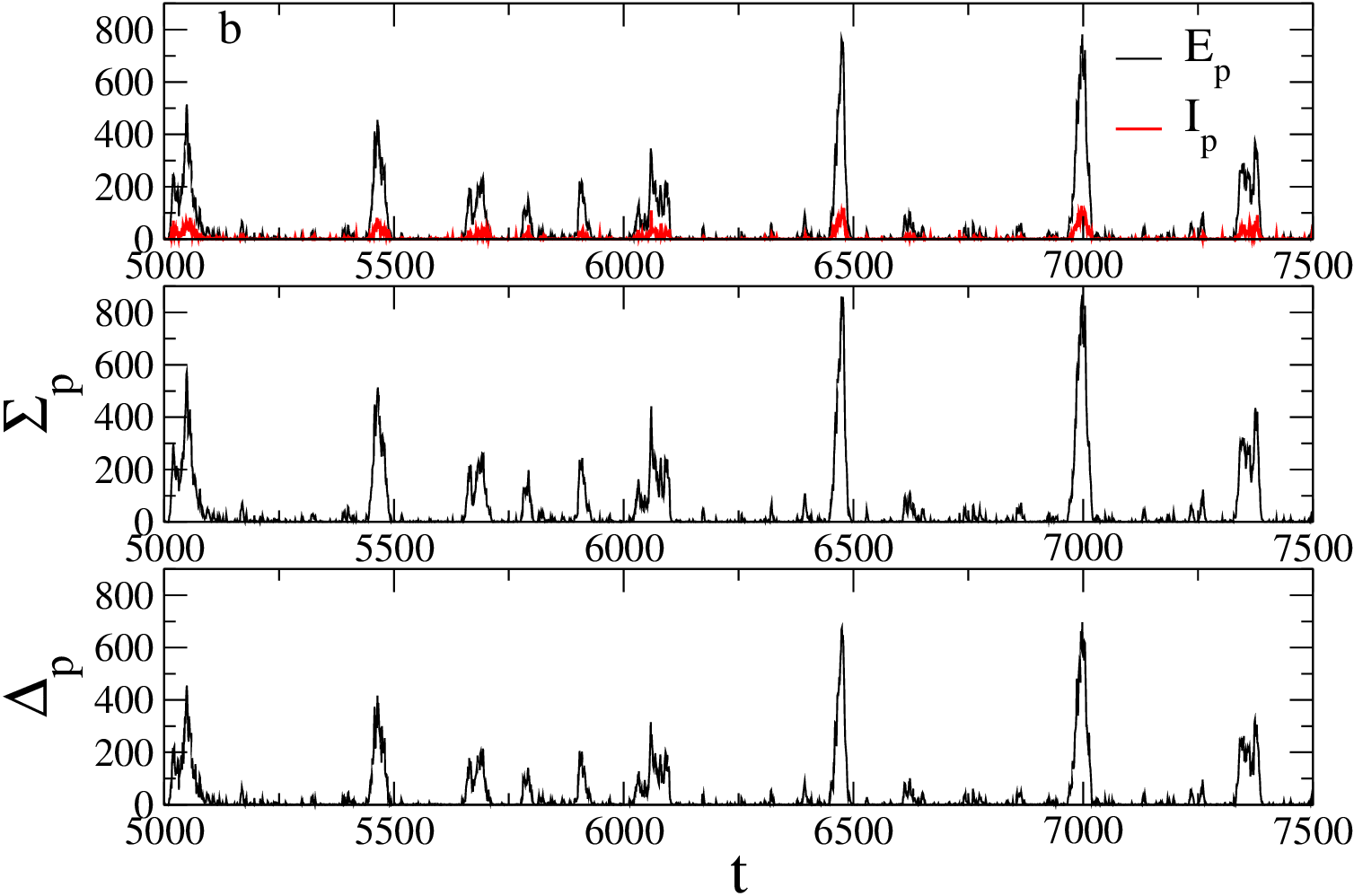}
	\includegraphics[width=0.49\textwidth,clip=true]{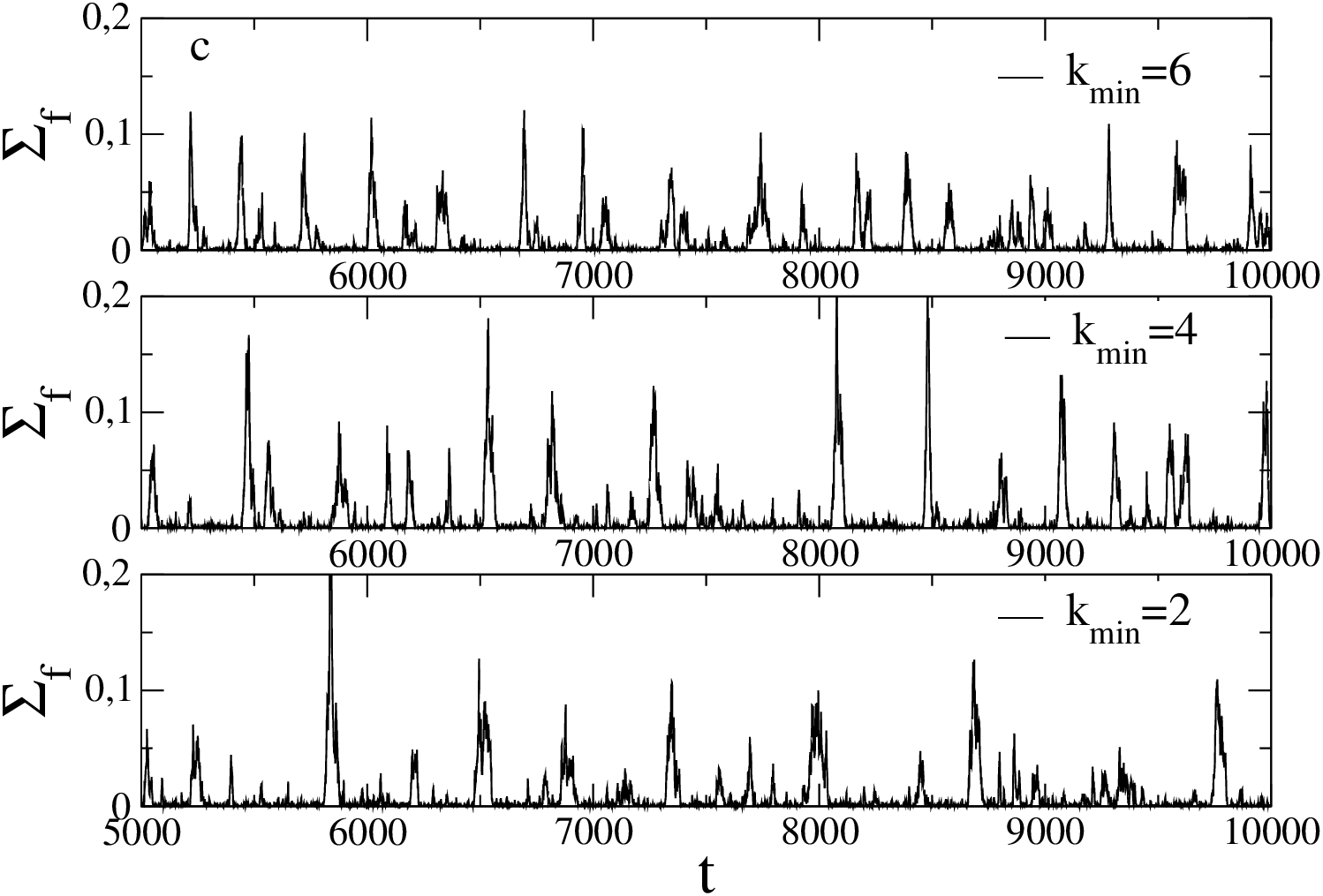}
	\includegraphics[width=0.49\textwidth,clip=true]{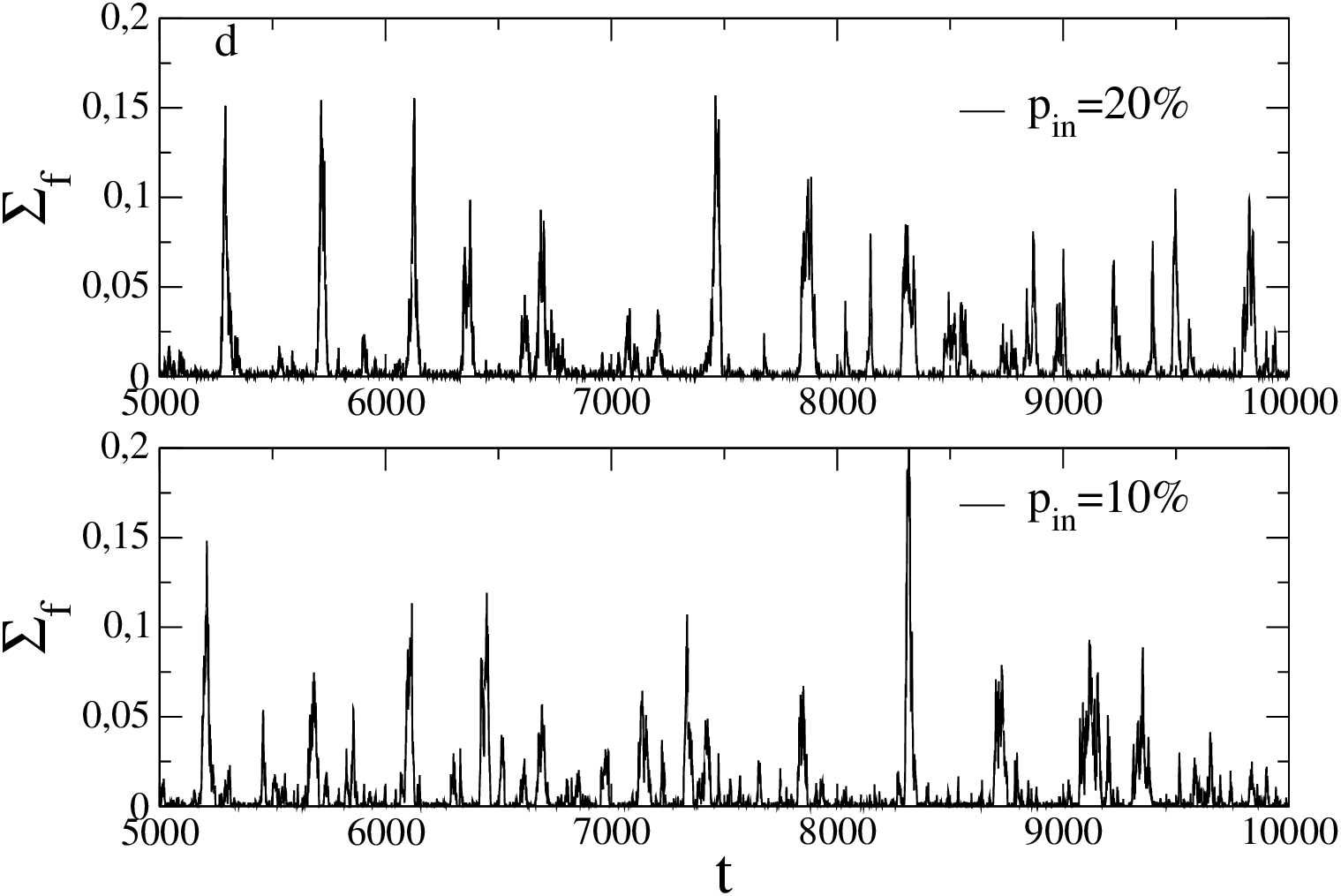}
	\caption{Temporal signals for the global activity $\Sigma$, the unbalance activity $\Delta$ and the activity of excitatory and inhibitory neurons, $E(t), I(t)$ in a network with $N=5000$ neurons with  $p_{in}=20\%$: $a)$ in the critical state with
		$\delta u_{recE}=0.00019$ and $\delta u_{recI}=0.000038$; $b)$ in the supercritical state with 	$\delta u_{recE}=0.00045$ and $\delta u_{recI}=0.00009$; $c)$ in the supercritical state for different values of $k_{min}$; $d)$ in the supercritical state for $p_{in}=10$ and 20\%. }
	\label{fig3}
\end{figure}

\section{Activity of excitatory and inhibitory neurons}
In order to monitor the activity in the different states, we analyse the temporal signal for the global activity $\Sigma (t)$ and for the activity of excitatory and inhibitory neurons separately, $E(t), I(t)$. Interesting information is also provided by the activity difference, $\Delta (t)=E(t)-I(t)$, which enlightens the unbalance in activity expected in the supercritical state. Activity can be measured, both, in terms of the number of firing neurons ($\Sigma_f$) (Fig.3c-d) and the sum of potential variations $\delta v_j=v_j(t)-v_j(t-1)$ at post-synaptic neuron $j$ caused by the firing of the pre-synaptic neuron $i$ ($\Sigma_p$) (Fig.3a-b). These two different measures, evidence different behaviours. In the critical state (Fig.3a), all temporal signals exhibit the expected bump sequence evidencing the avalanche occurrence. However, in terms of the number of firing neurons (not shown) excitatory roughly equals inhibitory activity leading to a $\Delta$ signal fluctuating around zero, as found in \cite{gs20}. The same behaviour is found for activity in terms of potential variations if a single $\delta u_{rec}$ is implemented (not shown). However, for different recovery rates we observe that inhibitory activity is always smaller than the excitatory one, even for similar numbers of firing neurons, due to the limited resources of inhibitory neurons. Interestingly, in this last case we observe that in the supercritical regime the activity of excitatory neurons, as expected, largely overtakes the inhibitory one,
 as testified by the larger amplitude of the $\Delta$ signal (Fig.3b).
 
To better enlighten the role of inhibition we analyse the activity signal for different fractions (Fig.3d) and connectivity degree (Fig.3c) of inhibitory neurons. In the supercritical state, the global signal $\Sigma$ shows that bursts in activity tend to become more frequent in time and with smaller amplitude for increasing fractions $p_{in}$ and connectivity degree $k_{min}$ (Fig.3c,d). This observation confirms that inhibition has the dual role of controlling the activity amplitude and making avalanche occurrence more frequent:
Hindering the large avalanches limits the synaptic fatigue in the network favouring the onset of new avalanches. We stress that this feature can be evidenced only if activity is monitored in terms of potential variations, rather that number of firing neurons, since
this measure takes into account the intensity of the neuronal signal, namely the frequency in action potentials.

\section{Temporal correlation functions}
In order to better characterize the critical and supercritical state we evaluate the temporal auto-correlation function for the global signal
$$C_\Sigma=\overline{\frac{<\Sigma(t)\Sigma(0)>-<\Sigma(t)><\Sigma(0)>}{<\Sigma(0)\Sigma(0)>-
	<\Sigma(0)><\Sigma(0)>}} \eqno(4)$$ 
where the temporal average (brackets) is evaluated on consecutive windows of 1000 time steps (1 time step$\simeq 10$ms) and the ensemble average (overline) is on 8000 network configurations. The correlation $C_\Sigma$ is evaluated for both $\Sigma_f$ and $\Sigma_p$. Fig.4 (left) shows the different behaviour of $C_\Sigma$ in the subcritical, critical and supercritical states for the model with a single recovery rate.
In the subcritical state the correlation function decays to zero as a simple exponential. In the critical state $C_\Sigma$ exhibits an initial abrupt decay to small negative values, going to zero at long times, behaviour that can be well fitted by the difference between two exponentials $C(t) = A\exp(-t/\tau)-B\exp(-t /\tau_0)$.  Interestingly, the analysis of different system sizes (Fig.4 right) shows that in the critical regime the characteristic time of the exponential scales with $N$ as $\tau\sim \sqrt N$ suggesting that in the thermodynamic limit long-range temporal correlations emerge, as expected at criticality.

\begin{figure}[htbp]
	\centering
	\includegraphics[width=0.49\textwidth,clip=true]{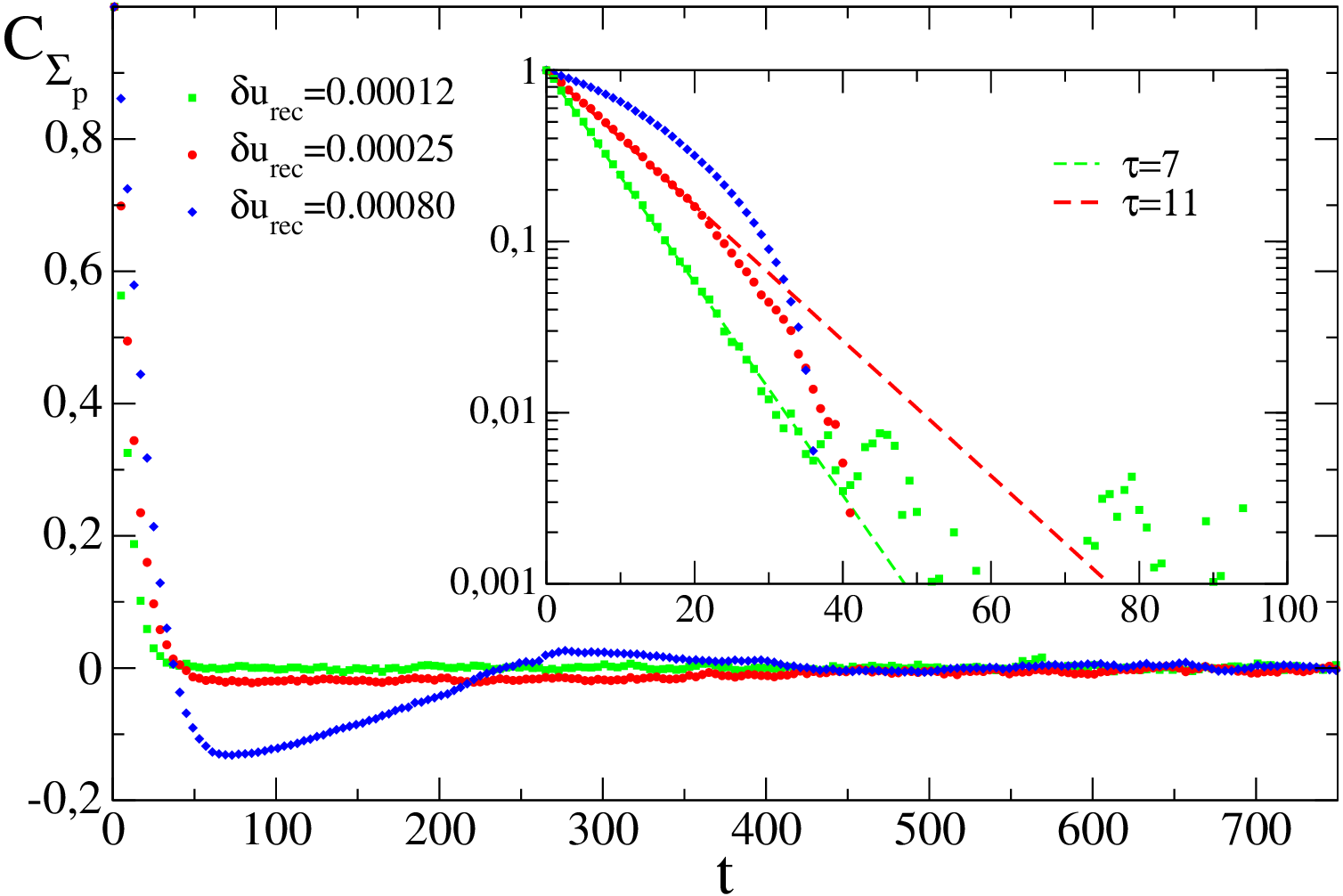}
	\includegraphics[width=0.49\textwidth,clip=true]{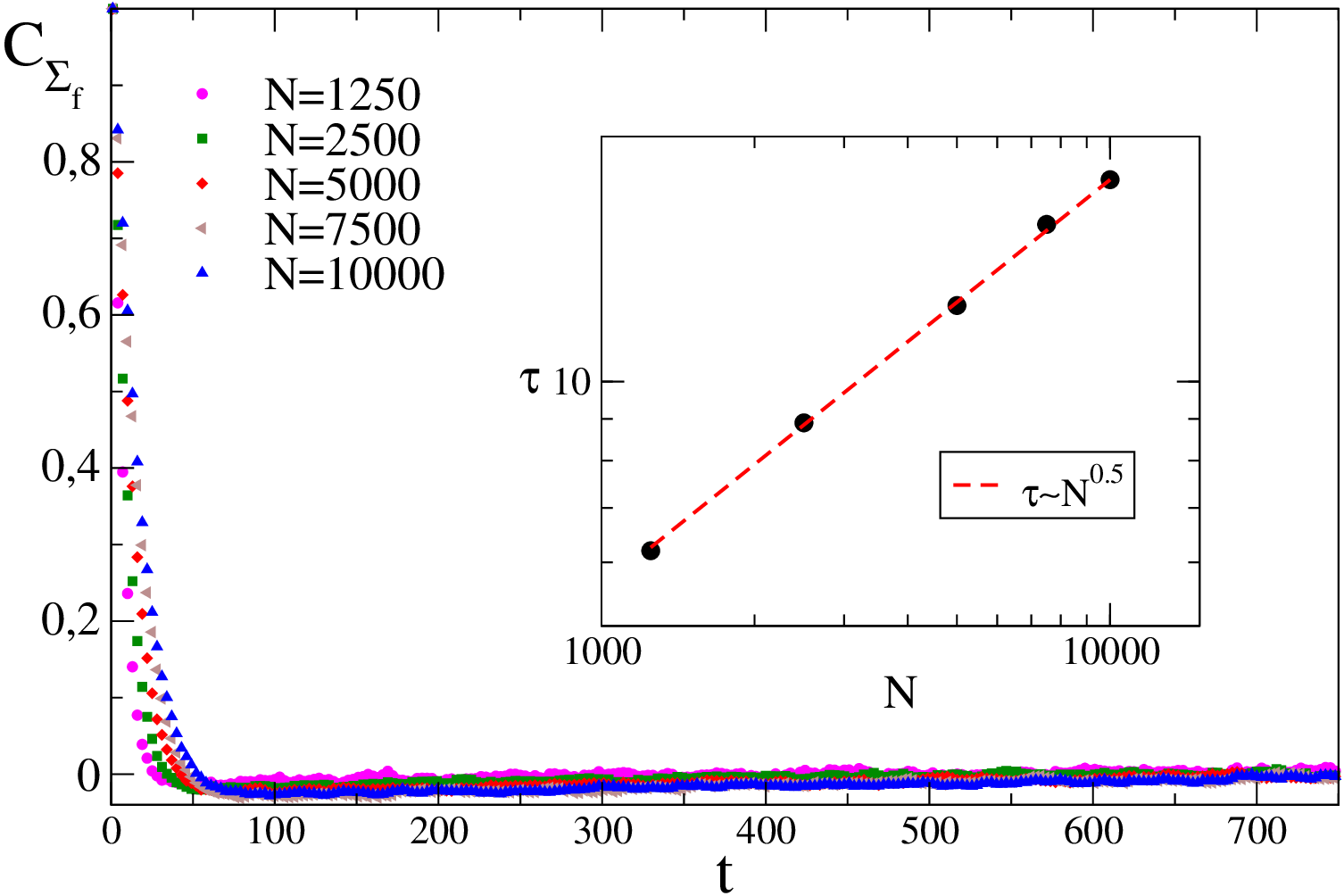}
	\caption{$Left$ Temporal correlation functions versus time for the global signal $\Sigma_p$ in terms of potential variations in the subcritical, critical and supercritical state for increasing $\delta u_{rec}$. The inset shows the same data in semi-log scale. $Right$ Autocorrelation function for the global activity in terms of the number of firing neurons $\Sigma_f$ in the critical state for different system sizes $N$ and $p_{in}=20\%$. The inset shows the characteristic time $\tau$ vs.$N$ on a log-log scale.}
    \label{fig4}
\end{figure}

Finally, in the supercritical state, the initial exponential decay is followed by a smooth oscillation going to zero at long times. This functional behaviour is more complex and can be fitted by $C(t) = [A\exp(-t/\tau_1) +(1-A)\exp(-t /\tau_2) ] \cdot [\cos(2\pi f t) + C\sin(2\pi f t)]$, where $f$ is a measure of the frequency of the observed oscillations. In Fig.5 we analyse the dependence of the fit parameters on the fraction $p_{in}$ and the connectivity degree $k_{min}$ of inhibitory neurons. Whereas the initial exponential decay is not affected by the change in parameters ($\tau_1\simeq 10$), both $\tau_2$ and $f$ change accordingly for an increasing influence of inhibitory neurons. In particular, both the period of oscillations and the characteristic time of the second exponential decrease as the fraction and the connectivity of inhibitory neurons increase.  

\begin{figure}[t]
	\centering
	\includegraphics[width=0.49\textwidth,clip=true]{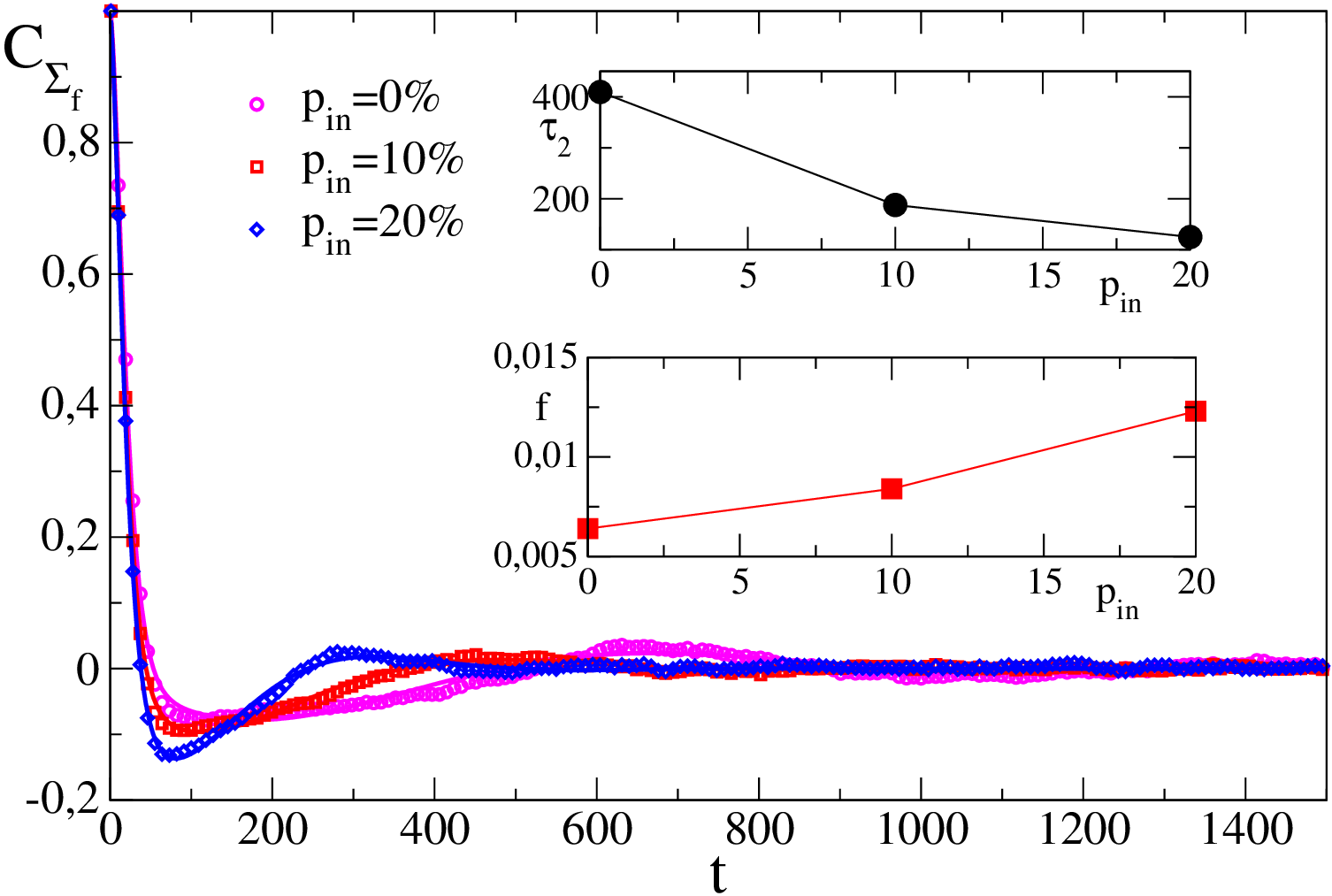}
	\includegraphics[width=0.49\textwidth,clip=true]{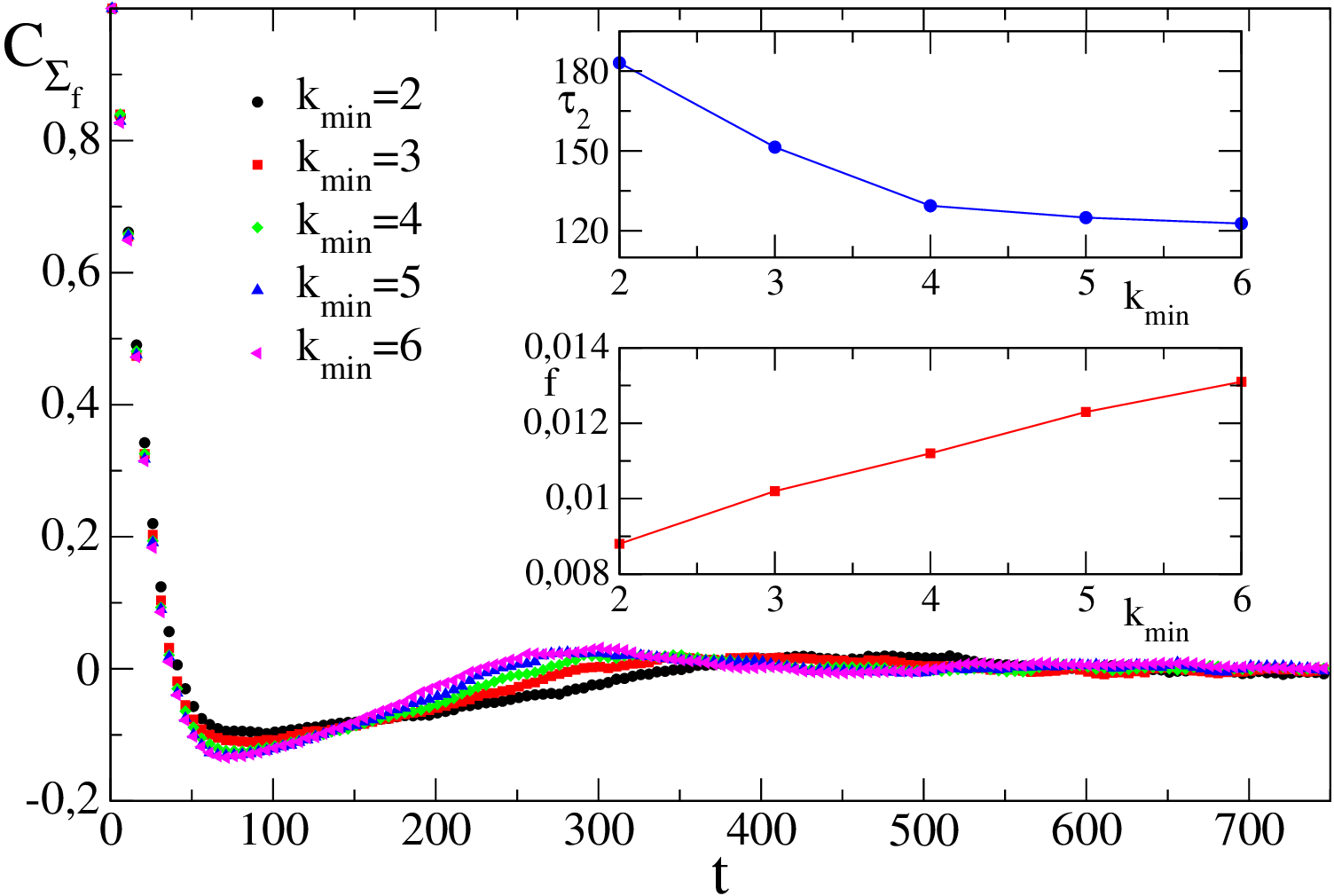}
	\caption{Temporal correlation functions versus time for the global signal $\Sigma_f$ in terms of number of firing neurons in the supercritical state of $N=5000$ networks ($left$) for different values of $p_{in}$ and ($right$) for $p_{in}=20\%$ and different $k_{min}$ values. The continuous lines show the fit with the function $C(t)$. The insets show $\tau_2$ and the frequency $f$ vs. $p_{in}$ and $k_{min}$. }
	\label{fig5}
\end{figure}

This is in agreement with the features of the temporal signal observed by varying the same parameters (Fig.3): Bursts in activity become more frequent and with a smaller amplitude as the relevance of inhibition in the network is increased. This effect is observed both increasing the percentage of inhibitory synapses and their average connectivity degree. The frequency of oscillations then appears to be related to the burst frequency in the temporal signal, whereas the faster decay of the second exponential could be related to the occurrence of smaller, and therefore shorter, avalanches. Increasing inhibition, in fact, shifts to the left the cutoff of the distribution, making avalanches as large as the system size not observed. As a consequence, the synaptic fatigue in the network becomes more localized, which increases the probability to trigger successive avalanches larger.

Finally, we analyse the correlation function for the model with a dual recovery rate (Fig.6). Whereas $C_\Sigma$ exhibits a similar behaviour to the model with a single rate in the critical state, the different role of inhibitory neurons modifies the correlation function in the supercritical state. In particular, even if the decay exhibits the same functional form,the frequency $f$ decreases if the dual recovery rate is implemented. This result can be rationalized by considering that the dual recovery rate limits the dissipative role of inhibitory neurons with respect to the model with a single recovery rate. 

\begin{figure}[t]
	\centering
	\includegraphics[width=0.49\textwidth,clip=true]{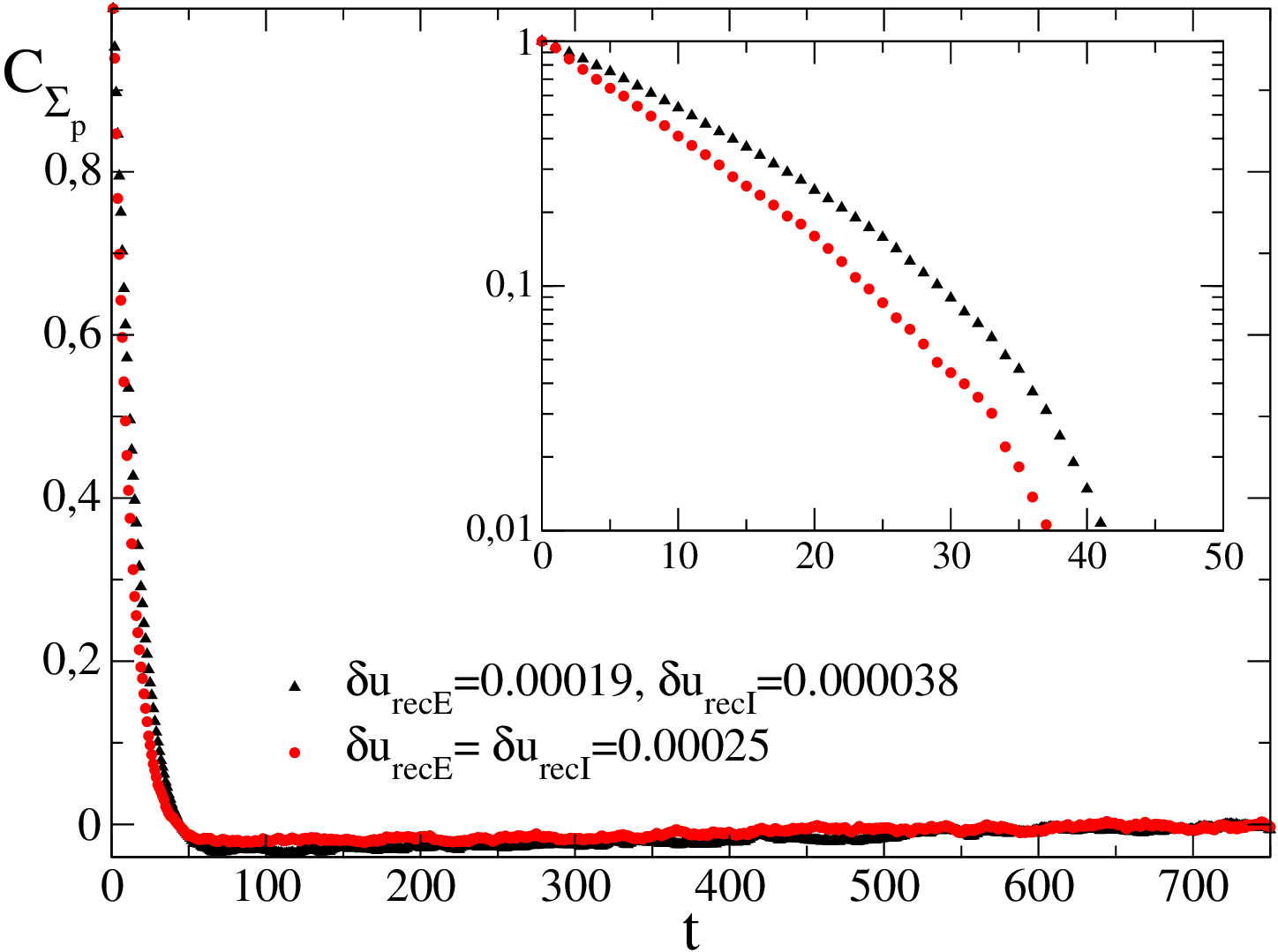}
	\includegraphics[width=0.49\textwidth,clip=true]{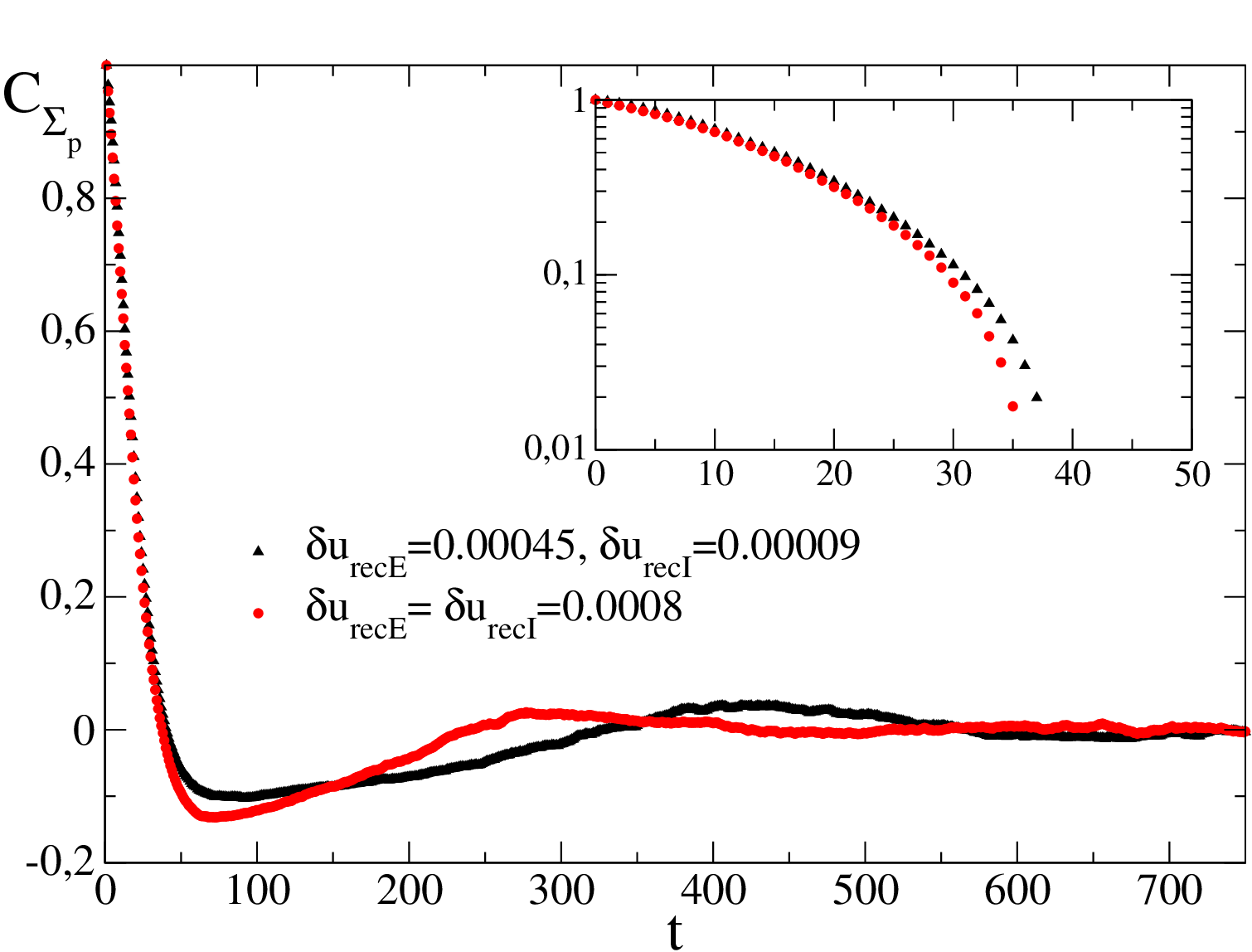}
	\caption{Temporal correlation functions versus time for the global signal $\Sigma_p$ in terms of potential variations in $N=5000$ networks with $p_{in}=20\%$ for
	the two cases of a single-valued and different $\delta u_{rec}$ in the critical ($left$) and supercritical ($right$) state. The insets show the same data in semi-log scale. }
	\label{fig6}
\end{figure}

\section{Conclusions}
Inhibitory neurons have a dissipative role in neuronal activity, which becomes more relevant as their outgoing connectivity degree increases. Indeed, the fact that inhibitory neurons are found to be hubs in the network suggests that, even if they are a minority, they are able to control activity over large regions. The analysis of the activity signal and its temporal correlation function confirms that critical vs. supercritical conditions are controlled by the different role of inhibition. A critical, healthy, activity is characterized by the balance of excitation and inhibition and a correlation function with an exponential decay, whose characteristic time diverges in the asymptotic limit. A more complex behaviour is found in the supercritical state: Unbalance of excitation with respect to inhibition is observed. The correlation function decays to negative values and goes to zero at long times with an oscillatory behaviour. The frequency of such oscillations depends on the fraction and the connectivity of inhibitory neurons. Indeed, a large inhibition hampers the occurrence of large avalanches, encompassing the entire system. Therefore, by keeping the activity more localized, the occurrence of successive avalanches in other regions of the network becomes more probable, increasing their frequency. Understanding the behaviour of correlations in spontaneous brain activity is extremely relevant since they could represent a proxy of different, healthy (critical) or pathological (supercritical), brain conditions.
Moreover, experimental evidence indicates that the ongoing state influences the response of the system performing a task \cite{arieli}. The fluctuation-dissipation relation, recently derived analytically in the framework of the Wilson-Cowan model \cite{fdt}, states that the response to an external stimulus is controlled by the correlations of activity fluctuations in the resting state. This result, confirmed by human magnetoencephalography data, can inspire new research lines on the controlled response of the brain.

\noindent
%\begin{acknowledgments}
{\it Acknowledgements.}	LdA would like to thank MIUR project PRIN2017WZFTZP for financial support.  LdA and AS acknowledge support from Program
	(VAnviteLli pEr la RicErca: VALERE) 2019 financed by the University
	of Campania ``L. Vanvitelli''. AS acknowledges support from MIUR project PRIN201798CZLJ.
%\end{acknowledgments}

%\section*{References}

\bibliography{ref}

\end{document}